
%
\input phyzzx


\font\largemath = cmti10 scaled \magstep2

\textfont4=\largemath
\mathchardef\lS="0453

\footline={\hfill}
\unnumberedchapters

\def\Gd{\delta}
\def\Ge{\epsilon}

\def\Gn{\nu}

\def\Gq{\theta}

\def\GF{\Phi}

\def\GP{\Pi}

\def\Pd{\partial}
\def\ol{\overline}

\def\overset#1\to#2{\mathop{#2}\limit^{#1}}
\def\underset#1\to#2{\mathop{#2}\limit_{#1}}
\rightline{{\bf\tt CHIBA-EP-70}}
\rightline{February, 1994}
\rightline{hep-th/9404026}
\vskip 1 cm
\date={}

\titlepage
\title{
Supersymmetry in Stochastic Quantization Method and
Field-Dependent Kernel}

\author{Kenji Ikegami}

\address{Graduate School of Science and Technology,  \break
                 Chiba University, \break
        1-33 ~Yayoi-cho,  ~Inage-ku, ~Chiba 263, JAPAN }%

\baselineskip = 12pt

\abstract{We define a discretized Langevin equation in
Stratonovich-{\it type} calculus. We show that a generating functional
with a field-dependent kernel can be written in mid-point prescription
only when we calculate in the calculus. Moreover we investigate whether
supersymmetry of the stochastic action with field-dependent kernel
exists or not. }%

\baselineskip = 18pt
%
\REF \parisi{G.Parisi and Y.Wu, Sci.Sin. {\bf 24}
(1981)483.}

\REF \damgaard{For a review, P.H.Damgaard and H.H\"uffel,
Phys.Rep.{\bf 152}(1987)227.}

\REF \ghost{M.Namiki, I.Ohba, K.Okano and Y.Yamanaka,
Prog. Theor. Phys. {\bf 69}\hfill\break (1983)1580.}

\REF \nakghost{A.Nakamura, Prog.Theor.Phys.
{\bf 86}(1991)925.}

\REF \regularization{For example, J.T.Breit, S.Gupta and A.Zaks,
Nucl.Phys.B233 (1984)61;\hfill\break
J.Alfaro, Nucl.Phys.B253(1985)464.}

\REF \fermi{U.Kakudo, Y.Taguchi, A.Tanaka and K.Yamamoto,
Prog.Theor.Phys.{\bf 69}\hfill\break
(1983)1225;\hfill\break
T.Fukai, H.Nakazato,I.Ohba,K.Okano and Y.Yamanaka, Prog.Theor.Phys.
{\bf 69}(1983)1600.}

\REF \gravity{H.Rumpf, Phys.Rev.D33(1986)942.}

\REF \ikemocyos{K.Ikegami, R.Mochizuki and K.Yoshida,
Prog.Theor.Phys.{\bf 89}(1993)197.}

\REF \tanaka{S.Tanaka, M.Namiki, I.Ohba, M.Mizutani, N.Komoike and
M.Kanenaga, \hfill\break Phys.Lett.B288(1992)129.}

\REF \ps{G.Parisi and N.Sourlas, Phys.Rev.Lett.43(1979)744;
\hfill\break Nucl.Phys.B206(1983)321.}

\REF \okano{H.Nakazato, K.Okano, L.Sch\"ulke and Y.Yamanaka,
Nucl.Phys.B346(1990)\hfill\break 611;\hfill\break
S.Marculescu, K.Okano and L.Sch\"ulke, Nucl.Phys.B349(1991),463.}

\REF \mockernel{R.Mochizuki, Prog.Theor.Phys. {\bf 85}(1991)407.}

\REF \super{E.Egorian and S.Kalitzin, Phys.Lett.B129(1983)320;
\hfill\break
R.Kirschner, Phys.Lett.B139(1984)180;\hfill\break
E.Gozzi, Phys.Lett.B143(1984)183.}

\REF \ito{K.Ito, Proc.Imp.Acad. {\bf 20} (1944)519.}

\REF \stra{R.L.Stratonovich, Conditional Markov Processes and
Their Application to the Theory of Optimal Control(Elsevier,
NewYork, 1968).}

\REF \Gozzi{E.Gozzi, Phys. Rev. D28 (1983) 1922.}

\REF \SQ{For example, M.Namiki, Stochastic Quantization
(Springer-Verlag, 1992)\break p.209.}

\REF \ezawa{H.Ezawa and J.R.Klauder, Prog. Theor. Phys. 74(1985)904.}
%
\footline={\hfill-- \folio\ --\hfill}
\pagenumber = 1

\section{Introduction}
 The stochastic quantization method (SQM) was first
proposed by Parisi and
Wu as an alternative quantization method in 1981.
\refmark\parisi \refmark\damgaard
SQM can be applied to gauge theories without the
gauge fixing procedure, i.e. without Faddeev-Popov ghost
fields. Instead of introducing ghost field, the method
produces the same contribution as the path-integral quantization
method (PIQM). This fact was already confirmed perturbatively for
Yang-Mills fields\refmark\ghost
and for non-Abelian anti-symmetric tensor fields.
\refmark\nakghost

SQM has a powerful tool, `` kernel'', which, among others,
gives new regularization schemes.\refmark\regularization
Kernel is also introduced for systems including massless fermion.
\refmark\fermi
Moreover, ``field-dependent'' kernel is introduced for systems
including  graviton,\refmark\gravity systems with spontaneously
broken symmetry,\refmark\ikemocyos
and bottomless systems.\refmark\tanaka
 On the other hand, it is well known that theories quantized
stochastically display \break supersymmetry.\refmark\ps\refmark\okano
So my question is whether SQM with {\it field-dependent} kernel has
supersymmetry or not. In this paper we show that SQM with
field-dependent kernel has supersymmetry as well as the one without
kernel. While Ref.[\mockernel] showed that stochastic action with
field-dependent kernel was invariant by operation with the
supersymmetry generator\footnote\sharp{The definition of
$Q,\overline{Q}$ are given in this paper}$Q$, it did not show that the
action was invariant by operation with
$\overline{Q}$ and could be described in superfield
formalism.\refmark\super  Besides, boundary condition\refmark\okano of
generating functionals function was not discussed in Ref.[\mockernel].

When we construct the stochastic action, Ito\refmark\ito  or
Stratonovich\refmark\stra calculus can be used. Leibnitz rule, which is
indispensable to supersymmetry, can be used in the Stratonovich
calculus, but cannot be used in Ito calculus in continuum limit.
Therefore, we define a discretized Langevin equation in
Stratonovich-{\it type} calculus, and we construct the stochastic
action where Leibnitz rule can be used. In this paper, we show that
only in the calculus the stochastic action in {\it mid-point
prescription} can be constructed.


\section{Boundary condition of generating functional}

First, we take up a system with variables $q(x)$ and
the classical action $S(q)$ in {\it n}-dimensional space-time.
To quantize the system stochastically,
we consider the fictitious time interval $[-T,T]$ and divide the
interval to $2N$ segments with space $T/N $. We shall let T tend
to infinity later.
Besides, we define the discretized Langevin equation with the
field-dependent kernel $K(\ol{q}_i)$
$$
\eqalign{dq_i(x) &\equiv q_{i+1} - q_{i}
            = -K(\ol{q}_i){\Gd S \over \Gd \ol{q}_i}
    + R_i{\Gd R_i\over \Gd \ol{q}_i} +R_i dW_i ,\ \
\cr
       i=1,2,&\cdots,N,\ \ \ \ \ -T=t_{-N}<t_{-N+1}<\cdots <t_N=T,
\cr
        q_i(x) &\equiv q(x,t_i), \ \ \ \ \
            \ol{q}_i\equiv {1\over 2}(q_{i+1}+q_{i}),\ \ \ \ \
        R_i\equiv R(\ol{q}_i)
\cr
        K(\ol{q}_i)&= R^2(\ol{q}_i), \ \ \ \ \ \
             dW_i(x)\equiv W(x,t_{i+1}) - W(x,t_{i}),} \eqno(1)
$$
where $dt=T/N$ and we assume the kernel $K(\ol{q_i})$ is
positive-definite. $W(x,t_i)$ is a Wiener process defined by the
following correlation
$$
\eqalign{\langle (W(x,t_i)-W(x,t_k))&(W(y,t_j)-W(y,t_l)) \rangle
                    = 2(t_j-t_k)\Gd^n (x-y),\cr
        &i>j>k>l,\cr
\langle \cdots \rangle = &\int D(dWi) (\cdots ) exp[-{1\over 4}
\smallint d^nx\sum_{i=-N}^N ({dW_i\over dt})^2 dt].}\eqno(2)
$$
The expression of eq.(1) is different from the ordinary one in
Stratonovich calculus\refmark\SQ or the one in Stratonovich-related
calculus,\refmark\ezawa but we want to regard it as the expression in
Stratonovich-{\it type} calculus. The advantage of the calculus is
shown later.

Now we are able to show that Leibnitz rule can be used
in the Stratonovich-type calculus in the continuum limit
$dt\rightarrow 0$ as follows.
We shall calculate the expectation value
$$
\eqalign{
\smallint d^ny\langle &{\Gd f(\ol{q}_i)\over \Gd \ol{q}_i(y)} dq_i(y)
 \rangle
\cr
 &=
\smallint d^ny \langle {\Gd f(\ol{q}_i)\over \Gd \ol{q}_i(y)}\{
(-K(\ol{q}_i){\Gd S(\ol{q}_i)\over\Gd \ol{q}_i(y)}
+ R_i {\Gd R_i \over \Gd\ol{q}_i (y)} ) dt
+ R_i dW_i(y) \} \rangle,
\cr
&=
\smallint d^ny \langle {\Gd f(\ol{q}_i)\over \Gd \ol{q}_i(y)}
(-K(\ol{q}_i){\Gd S(\ol{q}_i)\over\Gd \ol{q}_i(y)}
+ 2R_i {\Gd R_i \over \Gd\ol{q}_i(y)} ) dt
+ K(\ol{q}_i){\Gd^2 f(\ol{q}_i)\over \Gd \ol{q}_i(y)^2} dt \rangle ,}
\eqno(3)
$$
where $f(\ol{q}_i)$ is some function of $\ol{q}_i$. In second equality
we use the fact that \break
$\langle q_i\ dW_i\rangle =0,\ \langle q_{i+1}\ dW_i\rangle =
\langle 2R(\ol{q}_i)dt\rangle +O(dt^2)$.
The final expression in eq.(3) is equivalent to the value
$$
  \smallint d^ny \langle {\Gd f(q_{i})\over \Gd q_{i}(y)}
(-K(q_{i}){\Gd S(q_{i})\over\Gd q_{i}(y)}
+ 2R(q_{i}){\Gd R(q_{i})\over \Gd q_{i}(y)} )dt
+ K(q_{i}){\Gd^2 f(q_{i})\over \Gd q_{i}^2(y)})dt \rangle, \eqno(4)
$$
up to $dt$ because $\ol{q}_i=q_i+{1\over 2}dq_i$. The expression
coincides with \break
$\langle df(q_i)\rangle \equiv \langle f(q_{i+1})-f(q_i)\rangle$
in Ito calculus because the Langevin equation in Ito calculus
is
$$
dq_i= -K(q_i){\Gd S(q_i)\over \Gd q_i}dt
+ 2R(q_i){\Gd R(q_i)\over \Gd q_i}dt
+R(q_i)dW_i.\eqno(5)
$$
Eqs.(3) and (4) mean
$$
\langle df(q_i)\rangle =\smallint d^ny
\langle {\Gd f(\ol{q}_i)\over \Gd \ol{q}_i(y)}dq_i(y) \rangle
+O(dt^2).\eqno(6)
$$
Therefore, Leibnitz rule can be used in continuum limit of the
Stratonovich-type calculus.

Next, let us introduce the generating functional
$$
\eqalign{Z[J]&\equiv
      \int D(dW_i)\ exp\Bigl[ -\smallint\ d^nx
           \sum_{i=-N}^N \{ {1\over 4}
     ({dW_i\over dt})^2+ J_i(x)q^W_i(x)\}dt \Bigr],\cr
   &=\int Dq_i\ det(M_{ij})\cr
  \times exp&\Bigl[ -\smallint d^nx \sum_{i=-N}^N[
     {1\over 4} \{ ({dq_i\over dt} + K{\Gd S(q_i)\over \Gd \ol{q}_i}
     -R_i{\Gd R_i \over \Gd \ol{q}_i})R_i^{-1}\}^2
   +J_i(x)q^W_i(x)] dt\Bigr],\cr
\ \ \ \ \ M_{ij}&\equiv {\Gd (dW_i)\over \Gd q_{j+1}},}\eqno(7)
$$
where $q^W_i$ is a solution of eq.(1).
Another expression of the generating functional is
$$
\eqalign{&\tilde{Z}[J] =
       \int  (DqDpD\ol{C } DC )\ e^{-A+\smallint d^nxdt J(x,t) q(x,t)},
  \cr
  & A\equiv \smallint d^nx \smallint^T_{-T}dt
        \{ pK(q)p
     -ip(\dot{q}+K{\Gd S\over\Gd q}-R{\Gd R\over\Gd q}) \cr
  &\ \ \ \ \ \ \ \ \ \ \ \ \ \   -\overline{C} R(q){\Pd\over\Pd q}
       R^{-1}(\dot{q}+K{\Gd S\over\Gd q} - R{\Gd R\over\Gd q} )C\},
          }\eqno(8)
$$
in continuum limit. Here the auxiliary field $p$ and Grassmannian
variables $\overline{C },C$ are introduced.  In order
to assert that $Z=\tilde{Z}$, we need to prove
$$
{\rm \mathop{lim}_{dt\rightarrow 0}}det(M_{ij})= I_{fermi},\eqno(9)
$$
where
$$
\eqalign{I_{fermi}&\equiv \int DpDCD\ol{C}\  e^{-A_f},\cr
A_f\equiv \smallint d^nx\smallint^T_{-T}dt[
\{ ( p-{i\over 2}&(\dot{q}
 +K{\Gd S\over\Gd q}-R{\Gd R\over\Gd q})K^{-1}) R \}^2\cr
&\ \ \ \ \ \ \  -\overline{C} R(q){\Pd\over\Pd q}
      R^{-1}(\dot{q}+K{\Gd S\over\Gd q} - R{\Gd R\over\Gd q} )C ]
.}\eqno(10)
$$
In order to calculate $det(M_{ij})$ and $I_{fermi}$, we take the
twisted boundary condition\refmark\okano
$$
q(-T)=e^{-i\Gn}q(T),\   C(-T)=e^{-i\Gn}C(T),\   e^{-i\Gn}\ol{C}(-T)=
 \ol{C}(T)
,\   e^{-i\Gn}p(-T)=p(T).\eqno(11)
$$
The boundary condition for $\Gn =0$ corresponds to the periodic one
$$
q(-T)=q(T),\ \ C(-T)=C(T),\ \ \ol{C}(-T)=\ol{C}(T),\ \ p(-T)=p(T),
\eqno(12)
$$
or for $\Gn =-i\infty$ the causal and anti-causal one
$$
q(-T)=0,\ \  C(-T)=0,\ \  \ol{C}(T)=0,\ \ p(T)=0,\eqno(13)
$$
In this paper, we choose $\Gn $ as $Re(\Gn )=0$ because
$q$ is a real variable. We can calculate $det(M_{ij})$ as
$$
\eqalign{&det(M_{ij})=\mathop{\GP}_{i=-N}^N R_i^{-1}  \cr
&\times det\left( \matrix{
 1+G_{-N}dt     &\hfil           &\hfil           &\hfil           &
e^{-i\Gn}(-1+G_{-N}dt)       \cr
-1+G_{-N+1}dt   & 1+G_{-N+1}dt   &\hfil           &\hfil           &
\hfil                        \cr
\hfil           & \hfil          &\hfil           &\hfil           &
\hfil                        \cr
\hfil           &\hfil           &\hfil           &\hfil           &
\hfil                        \cr
\hfil           &\hfil           &\hfil           & -1+G_N dt      &
\hfil  1+G_N dt              \cr
}\right),\cr
&\ \ \ \ \ G_i dt\equiv
{1\over 2}R_i{\Gd R_i^{-1}\over\Gd\ol{q}_i}(q_{i+1}-q_i)
 + {1\over 2}R_i{\Pd\over\Pd\ol{q}_i}
   (R_i{\Gd S(\ol{q}_i)\over\Gd\ol{q}_i}
    -{\Gd R_i \over\Gd\ol{q}_i})dt.
}\eqno(14)
$$
The Stratonovich-{\it type} calculus has the advantage that the
$det(M_{ij})$ is expressed in the mid-point prescreption, i.e.,
$$
det(M_{ij})=\mathop{\GP }_{i=-N}^{N}
   g({q_{i+1}+q_i\over 2},q_{i+1}-q_i),
\eqno(15)
$$
where $g$ is some function. In ordinary Stratonovich calculus,
$det(M_{ij})$ cannot be expressed in the prescreption in case that
kernel is field-dependent. By straightforwad calculation, the
determinant is
$$
\eqalign{\mathop{lim }_{dt\rightarrow 0 }det(M_{ij})&=
     \mathop{\GP }_{t=-T}^T R^{-1}(t) \ \
\{ e^{\smallint^T_{-T}dtd^nx\ G(t)}
            -e^{-\smallint^T_{-T}dtd^nx\ G(t)-i\Gn }\},\cr
&=\mathop{\GP }_{t} R^{-1}(t)\ {\rm sinh}
 \{ \smallint^T_{-T}dtd^nx\ G(t)
+{1\over 2}i\Gn\}\times {\rm Constant}.}\eqno(16)
$$
Next $I_{fermi}$ can be also calculated as
$$
I_{fermi}=
  {\rm sinh}[\{ {1\over 2}i\Gn + \smallint^T_{-T}d^nxdt\ G(t)\}].
\eqno(17)
$$
by following ref.[\Gozzi]. From eqs.(16) and
(17), we have proved that $Z$ is equivalent to $\tilde{Z}$ when the
boundary condition (11) is taken.

\section{Supersymmetry and field-dependent kernel}
{}From now on, we introduce the field-dependent kernel again. The
expression of $A$ in (8) is rather complicated and it is difficult to
recognize whether the stochastic action with field-dependent kernel has
supersymmetry or not.
 In fact, $A$ is not invariant under the supersymmetric transformations
under which the stochastic action without kernel is invariant.

Here, we consider the change of variables
$$
q'=\int dq R^{-1}(q), \ \ \ p'= R(q)p,\ \ \
\overline{C }'=\overline{C } R(q),
\ \ \ C'= R^{-1}(q)C.\eqno(18)
$$
This leads to
$$
\eqalign{&\tilde{Z}[J] = \int Dp'Dq' DC' D\overline{C }'
            e^{-A'+\smallint d^nxdt J(x,t)q(q')}\cr
  A'\equiv &\smallint d^nx \smallint^T_{-T}dt \{ p'^2 -ip'(\dot{q}'
        +{\Gd S \over \Gd q'}- R^{-1}{\Gd  R\over \Gd q'})
       -\overline{C }' {\Pd \over \Pd q'}
          (\dot{q}'+ {\Gd S \over \Gd q'}
   -R^{-1}{\Gd  R\over \Gd q'}) C'\}
             \Bigr],}\eqno(19)
$$
where we assume that
the first relation in eq.(18) can be solved for $q$ in terms of $q'$.
The periodic boundary condition (12) can be also expressed in terms of
new variables as
$$
q'(-T)= q'(T),\ \ C'(-T)=C'(T),\ \  \ol{C}'(-T)=\ol{C}'(T),\ \
p'(-T)=p'(T),\eqno(20)
$$
i.e., periodic boundary condition. For $\Gn=-i\infty$, the boundary
condition (13) can be also expressed as
$$
q'(-T)=0,\ \ \ C'(-T)=0,\ \ \  \ol{C}'(T)=0,\ \ p'(T)=0\eqno(21)
$$
i.e., causal and anti-causal boundary condition, where we choose the
integral constant as $\int\ R^{-1}dq|_{q=0}=0$.

The stochastic action $A'$ with the boundary condition (20) is
invariant under the super-transformations
$$
\Gd q'=\overline{\Ge}C',\ \ \ \Gd \overline{C}'=-i\ol{\Ge}p',\ \ \
\Gd C'=0,\ \ \ \Gd p'=0 , \eqno(22)
$$
and
$$
\Gd q'= \ol{C}'\Ge,\ \ \ \Gd \ol{C}'=0,\ \ \ \Gd C'= -i\Ge p'- \Ge
\dot{q}',
\ \ \ \Gd p'=i\dot{\ol{C}}'\Ge .\eqno(23)
$$
On the other hand, $A'$ with (21) is not invariant under the
transformation (23) as well as the case of no kernel.\refmark\okano

In terms of original variables $q,C,\overline{C}$ and $p$, the
transformation (22) is expressed as
$$
\Gd q=\ol{\Ge}C,\ \ \
\Gd\ol{C}=\ol{\Ge}\ol{C}{\Pd R\over \Pd q}R^{-1}C-i\ol{\Ge}p,\ \ \
\Gd C=-R{\Pd R^{-1}\over \Pd q}\ol{\Ge}C,\ \ \
\Gd p =-R^{-1}{\Pd R\over \Pd q}\ol{\Ge}C p,\eqno(24)
$$
and the transformation (23) is
$$
\eqalign{&\Gd q= \overline{C}K(q)\Ge,
\ \ \ \Gd\overline{C}=0
           ,\cr
&     \Gd C=-i\Ge K(q)p -\Ge\dot{q}
           + \overline{C} \Ge {\Pd  R(q)\over \Pd q}R(q)C,
\cr
&\Gd p= i\dot{\overline{C}}\Ge
            +i\overline{C}{\Pd R(q)\over \Pd q}R^{-1}\dot{q}\Ge
            -\overline{C}\Ge {\Pd R(q)\over \Pd q}R(q) p
.}\eqno(25)
$$
Finally, we express the stochastic action in terms of the superfield
$\GF'$ as
$$
\eqalign{A'= -\smallint d^n x d^2\Gq \smallint^T_{-T}dt \{ &
       \overline{D}_{\Gq}\GF'D_{\overline{\Gq}}\GF'+ L(q(\GF'))
        -\Gd^n(0)\ln R(q(\GF'))\} ,
      \cr
 &D_{\ol{\Gq}}\equiv \Pd_{\ol{\Gq}}-\Gq\Pd_t,\ \
\ol{D}_{\Gq}\equiv \Pd_{\Gq},\ \
\GF'\equiv q' + \overline{\Gq}C' + \overline{C}'\Gq -i\overline{\Gq}
\Gq p',       }\eqno(26)
$$
where $\Gq,\ol{\Gq}$ are Grassmannian variables. As discussed above,
the expression is invariant under operation with
$Q(\equiv \Pd_{\ol{\Gq}})$ and
$\ol{Q}(\equiv \ol{\Pd}_{\Gq}+\ol{\Gq}\Pd_t)$
for
$\Gn=0$ and invariant under the operation with only $Q$ for
$\Gn=-i\infty$

\section{Summary}

 We defined the discretized Langevin equation (1) with field-dependent
kernel  in Stratonovich-{\it type} calculus in which Leibnitz rule can
be used and constructed the generating functional. When the
field-dependent kernel is introduced, only the generating functional
constructed in the Stratonovich-{\it type} calculus can be expressed
in the prescription. Besides, we showed that the generating functional
(7) constructed from Wiener process distribution is equivalent to the
continuous one (8) when we take the twisted boundary condition (11).

Moreover, we showed that the stochastic action with the periodic
boundary condition (12) is invariant under the super-transformations
(22) and (23) when field-dependent kernel is introduced. The stochastic
action with the causal and anti-causal boundary condition (13) is not
invariant under the super-transformation (23) as well as the one
without kernel.  We also show that the stochastic action can be also
described in terms of superfield.

\chapter{Acknowledgments}
 Author thanks Prof.S.Kawasaki, Drs.R.Mochizuki and A.Nakamura
for valuable discussion and comments.

\vfill\eject

\refout

\vfill\eject

\bye